\documentclass[aps,preprint]{revtex4}

\usepackage{graphicx}

\newcommand{\be}{\begin{equation}}
\newcommand{\ee}{\end{equation}}

\begin{document}

\title{Lowest Order Constrained Variational calculation for
Polarized Liquid $^3\mathrm{He}$ at Finite Temperature }
\author{G.H. Bordbar\footnote{Corresponding author}
\footnote{E-Mail: bordbar@physics.susc.ac.ir}, M.J. Karimi and J.
Vahedi}
 \affiliation{ Department of Physics, Shiraz University, Shiraz 71454,
Iran}
\begin{abstract}
We have investigated some of the thermodynamic properties of spin
polarized liquid $^3\mathrm{He}$ at finite temperature using the
lowest order constrained variational method. For this system, the
free energy, entropy and pressure are calculated for different
values of the density, temperature and polarization. We have also
presented the dependence of specific heat, saturation density and
incompressibility on the temperature and polarization.
\end{abstract}

\maketitle

\section{Introduction}\label{S:intro}
The spin polarized liquid $^3\mathrm{He}$ at finite temperature is
an interesting many-body system which obeys the Fermi-Dirac
statistics. Several theoretical techniques and experimental method
have been used for investigating the properties of unpolarized
liquid $\mathrm{^3He}$ \cite{Wi,Ke,Vbf,Nd,Lt,Ty,Er,Sv}. Recently,
we have studied the unpolarized liquid $\mathrm{^3He}$ at finite
temperature and the spin polarized liquid $\mathrm{^3He}$ at zero
temperature \cite {Gh1,Gh2,Gh3}. In this calculations, we have
applied the lowest order constrained variational method based on
the cluster expansion of the energy functional. This method is a
powerful microscopic technique used in the many-body calculations
of dense matter \cite{Gh1,Gh2,Gh3,Gh4,Gh5,Gh6,Gh7,Gh8,Gh9,Gh10}.

In the present work, we use the lowest order constrained
variational method to investigate some of the thermodynamic
properties of spin polarized liquid $\mathrm{^3He}$ at finite
temperature. In our calculations, we employ the Lennard-Jones
\cite{LJ} and Aziz \cite{Aziz} potentials.

\section{Method\label{Met}}
We consider a system consisting of $N$ interacting $\mathrm{^3He}$
atoms with $N^{+}$ spin up and $N^{-}$ spin down atoms. These
atoms have a different chemical potential ($\mu$) and different
Fermi-Dirac distribution function \cite{Fet} as follows,
\begin{equation}
n^{\pm}(k)=\frac{1}{exp[\beta(\frac{\hbar^{2}k^{2}}{2m}-\mu^{\pm})]+1},
\end{equation}
where $\beta=\frac {1}{k_{B}T}$ and $T$ is the temperature. We
define the total number density by $\rho$ and the spin asymmetry
parameter by $\xi$ ($\xi=0$ is unpolarized and $\xi=1$ is fully
polarized cases),
\begin{eqnarray}
\rho&=&\rho^{+}+\rho^{-},\nonumber\\\xi&=&\frac{N^{+}-N^{-}}{N}.
\end{eqnarray}
Thus, to determine the chemical potential for every $\rho$, $T$
and $\xi$, we solve the following equation by numerical method,
\begin{equation}
\rho^{\pm}=
\frac{1}{2\pi^{2}}\int_{0}^{\infty}n^{\pm}(k)k^{2}dk\label{Jeq}.
\end{equation}
We calculate the energy using the lowest order constrained
variational method based on the cluster expansion of the energy.
We consider up to the two-body energy term in cluster expansion,
\begin{equation}
    E=E_{1}+E_{2}.
\end{equation}
For this system, the one-body energy per particle, $E_{1}$,  is
given by
\begin{eqnarray}
 E_{1}&=&E_ {1}^{+}+E_{1}^{-}\nonumber\\
 &=&\frac{\hbar^{2}}{4m\pi^{2}\rho}\left[\int_{0}^{\infty}n^{+}(k)k^{4}dk
 +\int_{0}^{\infty}n^{-}(k)k^{4}dk\right].
\end{eqnarray}
The two-body energy per particle, $E_{2}$, is defined
\begin{equation}
 E_{2}=\frac{1}{2N} \sum_{i,j}\langle ij\mid W(12) \mid
 ij-ji\rangle,
\end{equation}
where
\begin{equation}
W(r_{12})=\frac{\hbar^{2}}{m}(\nabla{f}(r_{12}))^{2}+f^{2}(r_{12})V(r_{12}).
\end{equation}
In the above equation, $V(r_{12})$ is the two body potential
between helium atoms  and $f(r_{12})$ is the two-body correlation
function. By considering the $|i\rangle$ as a plane wave, we have
derived the following relation for the two-body energy per
particle,
\begin{equation}\label{E_{2}}
 E_{2}=\frac{1}{2}\rho\int
 d\vec{r}\left[1-\frac{1}{4\pi^{4}\rho^{2}}\left[(\gamma^{+}(r))^{2}
 +(\gamma^{-}(r))^{2}\right]\right]W( r),
\end{equation}
where
\begin{equation}
   \gamma^{\pm}(r)=\int_{0}^{\infty}\frac{\sin
  (kr)}{kr}n^{\pm}(k)k^{2}dk.
\end{equation}
Now, we minimize the two-body energy Eq. (\ref{E_{2}}) with
respect to the variations in the two-body correlation function
subject to the normalization constraint \cite{Clark,Feen},
\begin{equation}
\frac{1}{N}\sum_{i,j}\langle ij\vert
h^{2}(r_{12})-f^{2}(r_{12})\vert ij-ji\rangle=1.
\end{equation}
The normalization constraint is conveniently re-written in the
integral form as
\begin{equation}\label{Cons}
 \rho \int
d\vec{r} [h^{2}(r)-f^{2}(r)]\Gamma(r)=1,
\end{equation}
 where
\begin{equation}
 \Gamma(r)=\left[1-\frac{1}{4\pi^{4}\rho^{2}}\left[(\gamma^{+}(r))^{2}
 +(\gamma^{-}(r))^{2}\right]\right].
\end{equation}
and $h(r)$ is the Pauli function,
\begin{equation}
 h(r) = [\Gamma(r)]^{-\frac{1}{2}}.
\end{equation}
The minimization of the two-body Energy $E_{2}$ gives the
following Euler-Lagrange differential equation for the two-body
correlation function
\begin{equation}\label{Dif}
f^{''}(r)+f^{'}(r)\frac{\Gamma^{'}(r)}{\Gamma(r)}-\frac{m}{\hbar^{2}}
\left(\lambda+V(r)\right)f(r) =0.
\end{equation}
The lagrange multiplier $\lambda$ imposes by normalization
constraint. The two-body correlation function calculated by
numerically integrating Eq. (\ref{Dif}) and then the energy per
particle of system can be obtained. Finally the Helmholtz free
energy per particle can be determined using the following
equation,
\begin{equation}
 F=E-TS.
\end{equation}
$S$ is the entropy per particle \cite{Fet} which is given by
\begin{equation}
   S=S^{+}+S^{-},
\end{equation}
where
\begin{equation}
  S^{\pm}=-\frac{1}{2\pi^{2}\rho}\int_{0}^{\infty}
  \left[n^{\pm}(k)\ln(n^{\pm}(k))
  +(1-n^{\pm}(k)\ln(1-n^{\pm}(k))\right]k^{2}dk.
\end{equation}
By calculating $E$, $S$ and $F$, we can obtain the thermodynamic
properties of the spin polarized liquid $\mathrm {^{3}He}$ at
finite temperature.

\section{Results\label{Res}}
\begin{figure}
\includegraphics[height=2.7in]{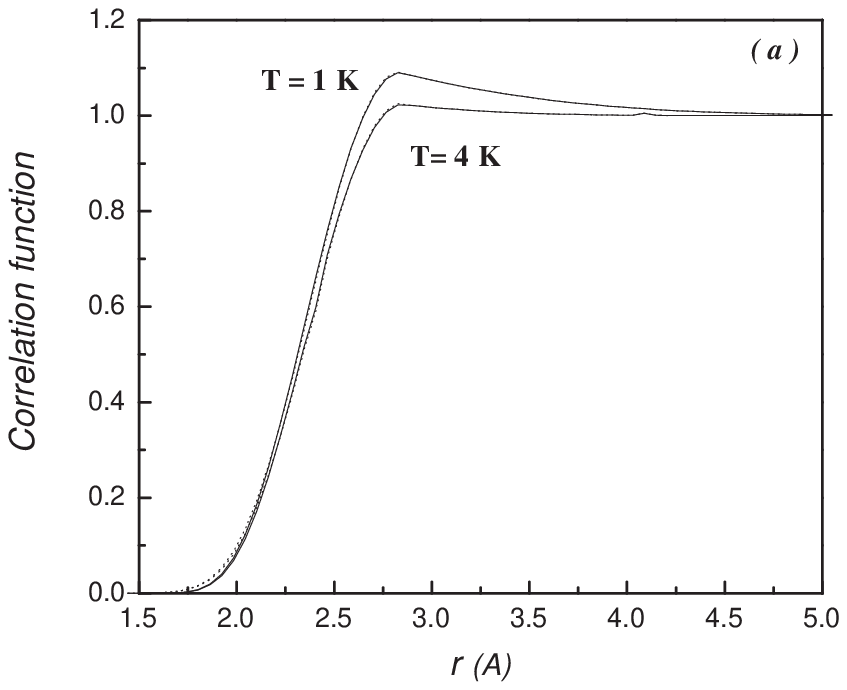}
\includegraphics[height=2.7in]{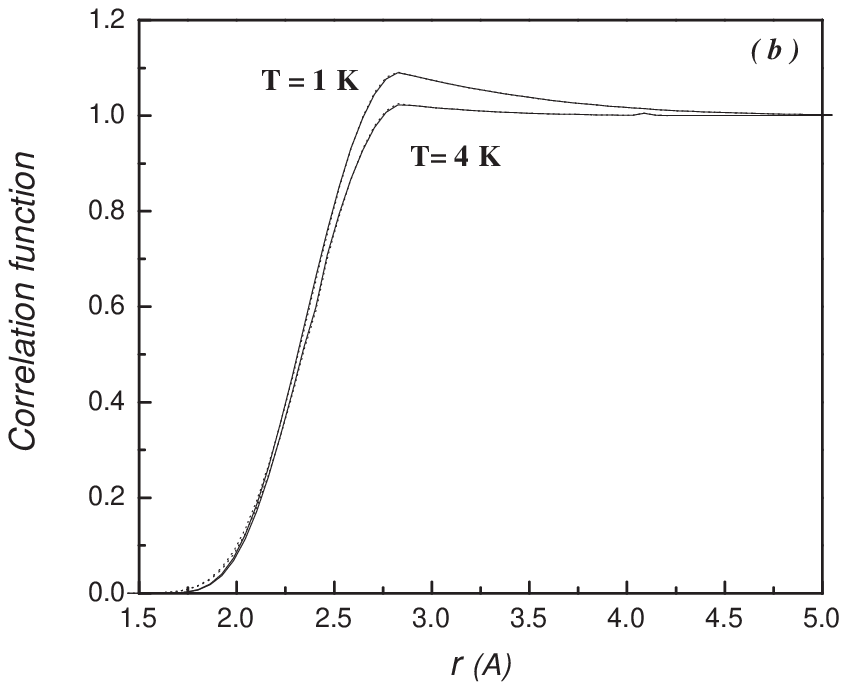}
\includegraphics[height=2.7in]{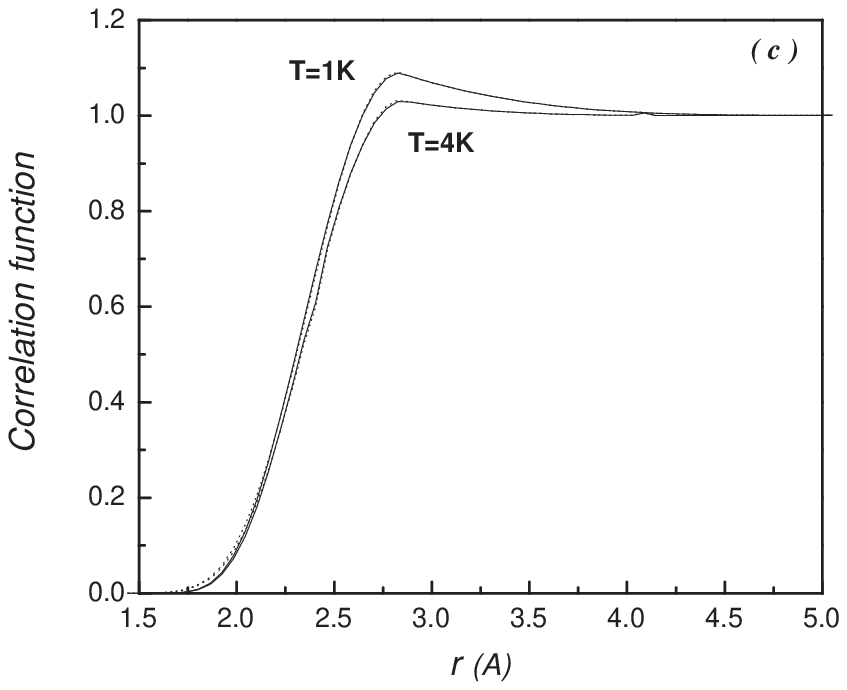}
\includegraphics[height=2.7in]{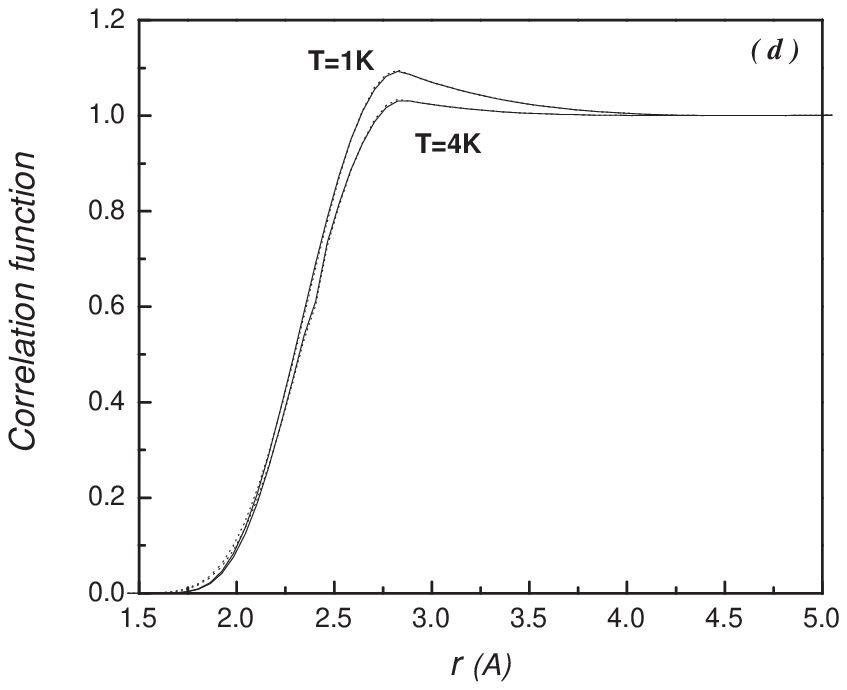}
 \caption{The correlation function versus the interatomic distance
($r$) with the Leonard-Jones (full curves) and Aziz (dotted
curves) potentials at $\rho=0.0166A^{-3}$ for  $\xi=0.0$ (a),
$\xi=0.4$ (b), $\xi=0.8$ (c) and $\xi=1.0$ (d).
  } \label{Cor}
\end{figure}

The two-body correlation function at $T=1.0 K$ and $T=4.0 K$ for
different values of the spin asymmetry parameter ($\xi$) are shown
in Fig. \ref{Cor}. This figure shows that at low temperatures, the
correlation function increases more rapidly and reaches the
limiting value ($f(r)=1$) at larger interatomic distances ($r$).
 It is shown that
by increasing $\xi$, the correlation function reaches the limiting
value at lower $r$. It is seen that our results for the
correlation function with the Lennard-Jones and Aziz potentials
are identical.

\begin{figure}
\includegraphics[height=2.7in]{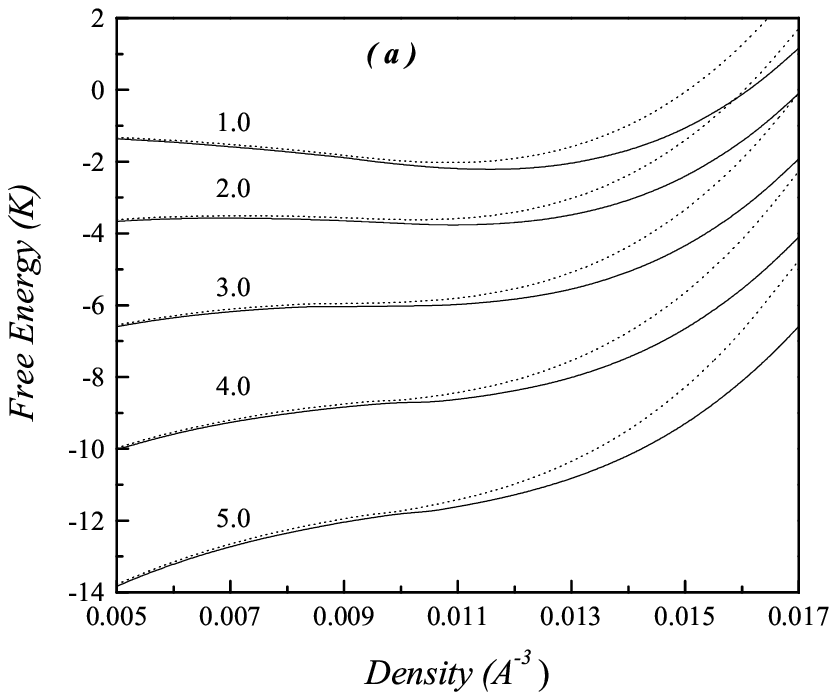}
\includegraphics[height=2.7in]{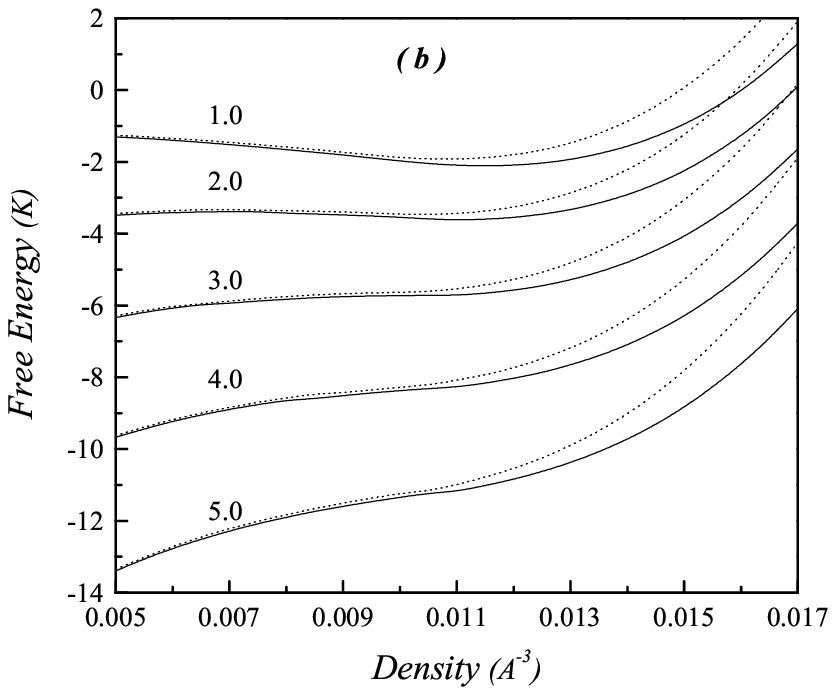}
\includegraphics[height=2.7in]{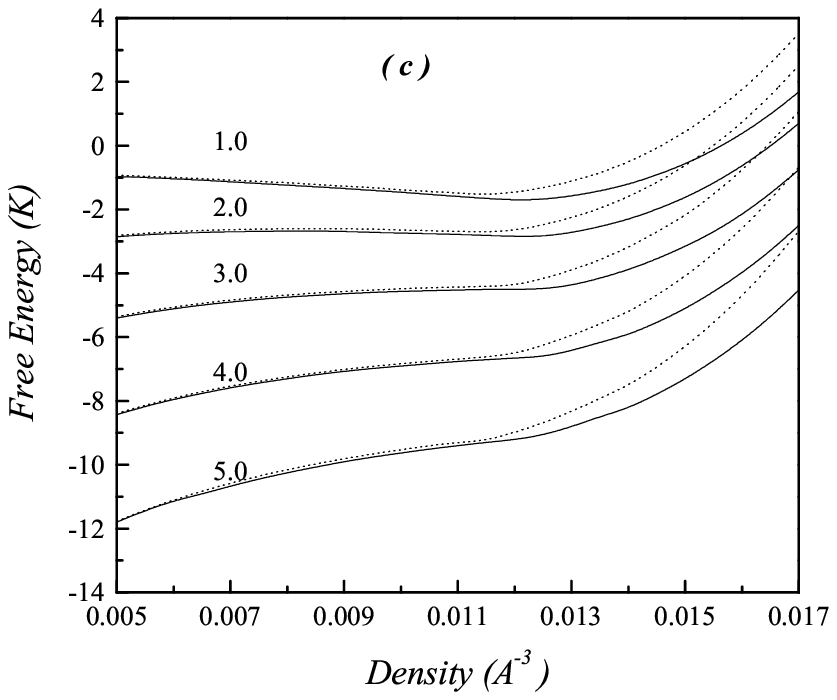}
\includegraphics[height=2.7in]{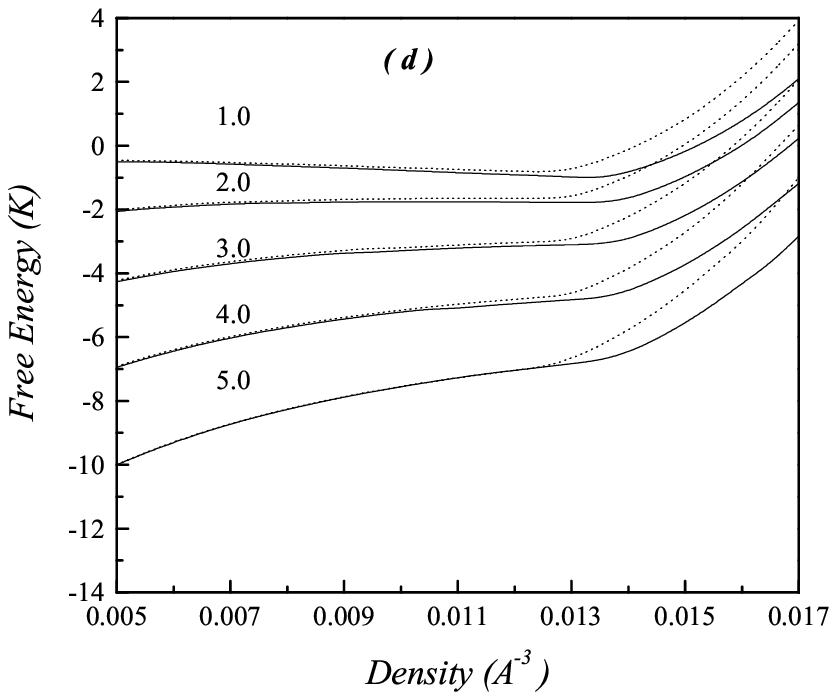}
 \caption{The free energy of liquid
$^3\mathrm{He}$ versus density at different values of temperature
($T=1.0, 2.0, 3.0, 4.0 , 5.0K$)with the Leonard-Jones (full
curves) and Aziz (dotted curves) potentials
 for  $\xi=0.0$ (a), $\xi=0.4$ (b), $\xi=0.8$ (c) and $\xi=1.0$
(d).
  } \label{freero}
\end{figure}

 Our results for the free energy of
liquid $\mathrm{^3He}$ with the Lennard-Jones and Aziz potentials
at different spin asymmetry parameters ($\xi$) are presented in
Fig. \ref{freero}.  We can see that at low densities ($\rho<
0.011A^{-3}$), the free energies with the Lennard-Jones and Aziz
potentials are similar. However, at higher densities, the
difference between these free energies are noticeable. Fig.
\ref{freero} shows that the free energy increases by increasing
spin asymmetry parameter and this variation becomes higher at high
temperatures. We see that the free energy doesn't show a bound
state above a certain value of temperature (called flashing
temperature, $T_{f}$ ). We have found $T_{f}$ for different values
of $\xi$ which given in Table \ref{Tab1}.  From Fig. \ref{freero},
it is also seen that for each value of spin asymmetry parameter,
the density of saturation point (minimum point of the free energy)
decreases by increasing temperature. We see that the saturation
density ($\rho_{0}$) increases by increasing spin asymmetry
parameter for each value of temperature.  Our results for the
saturation density are given in Table \ref{Tab2}. Table \ref{Tab1}
(Table \ref{Tab2}) shows that for all values of $\xi$ the flashing
temperature (the saturation density) with the Aziz potential is
lower than that of the Lennard-Jones potential.

\begin{figure}
\includegraphics[height=12cm]{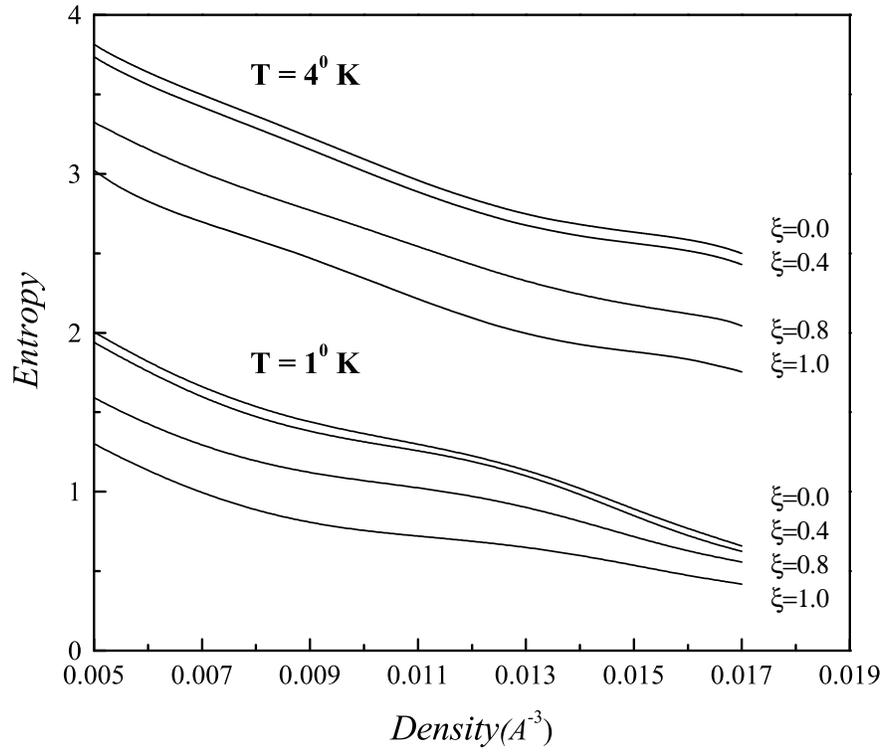}
 \caption{The entropy of liquid
$^3\mathrm{He}$ versus density at $T=1.0K$ and $T=4.0K$ for
different values of $\xi$. } \label{FigAnt}
\end{figure}

The entropy of liquid $\mathrm{^3He}$ is shown as a function of
density for two different temperatures and various asymmetry
parameters in Fig. \ref{FigAnt}. It is seen that the entropy
decreases by increasing both density and spin asymmetry parameter
and increases by increasing temperature. We have found that for
each temperature, the difference between entropy of fully
polarized and unpolarized liquid $\mathrm{^3He}$ decreases for
higher values of density, especially at low temperatures.

\begin{figure}
\includegraphics[height=12cm]{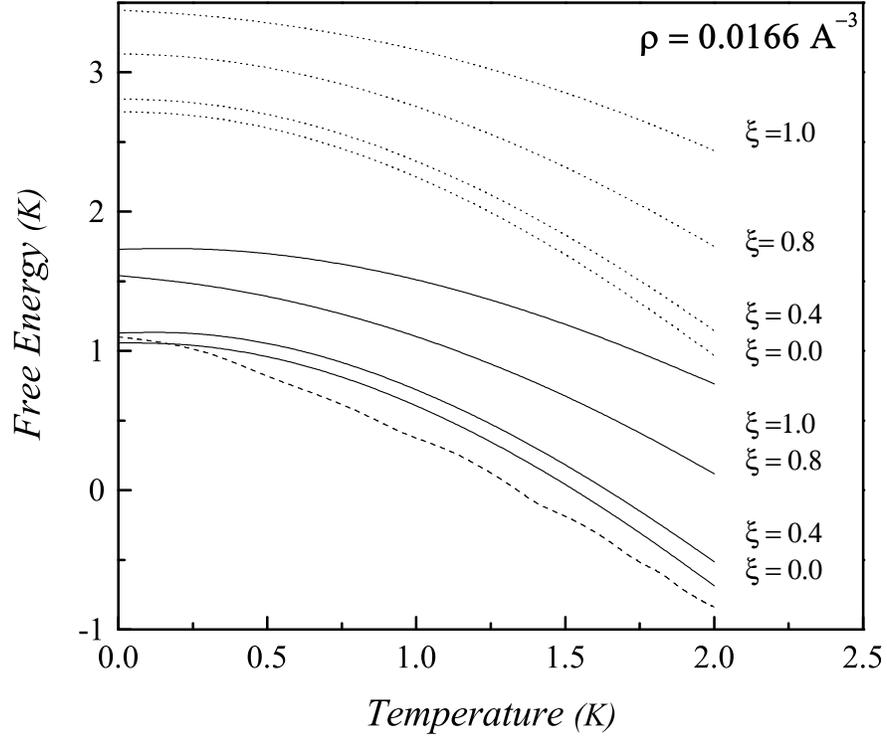}
 \caption{The free energy of liquid
$^3\mathrm{He}$ versus temperature for $\rho=0.0166A^{-3}$ at
different values of $\xi$ with the Leonard-Jones (full curves) and
Aziz (dotted curves) potentials. The experimental results (dashed
curve) for unpolarized liquid $^3\mathrm{He}$ \cite{Wi} are also
shown for comparison. } \label{FigFT}
\end{figure}

\begin{figure}
\includegraphics[height=12cm]{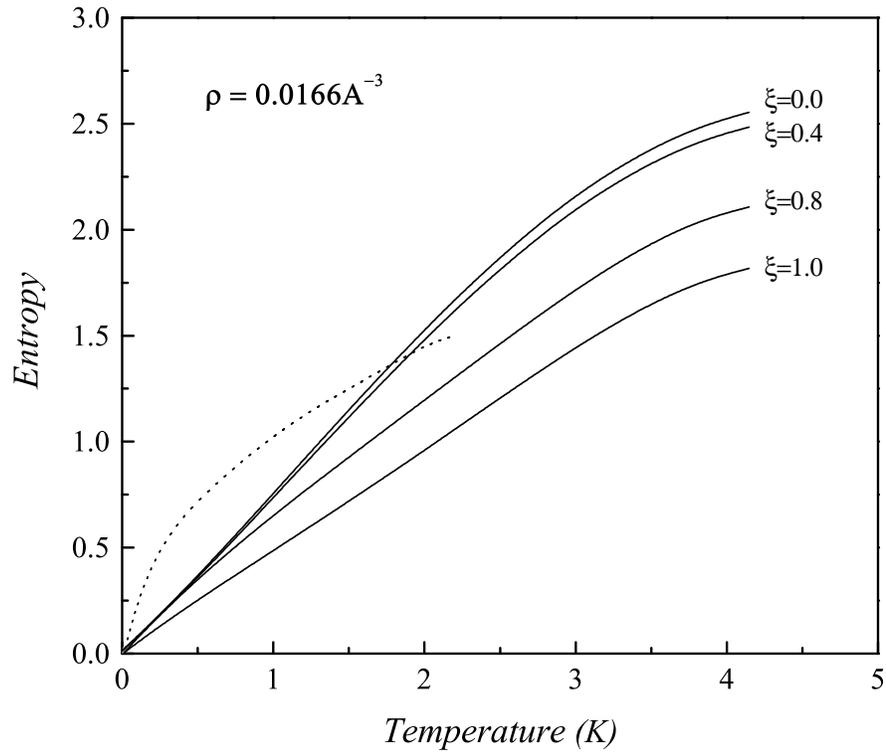}
 \caption{The entropy of liquid $^3\mathrm{He}$ versus temperature
for $\rho=0.0166A^{-3}$ at different values of $\xi$.}
\label{FigET}
\end{figure}

\begin{figure}
\includegraphics[height=12cm]{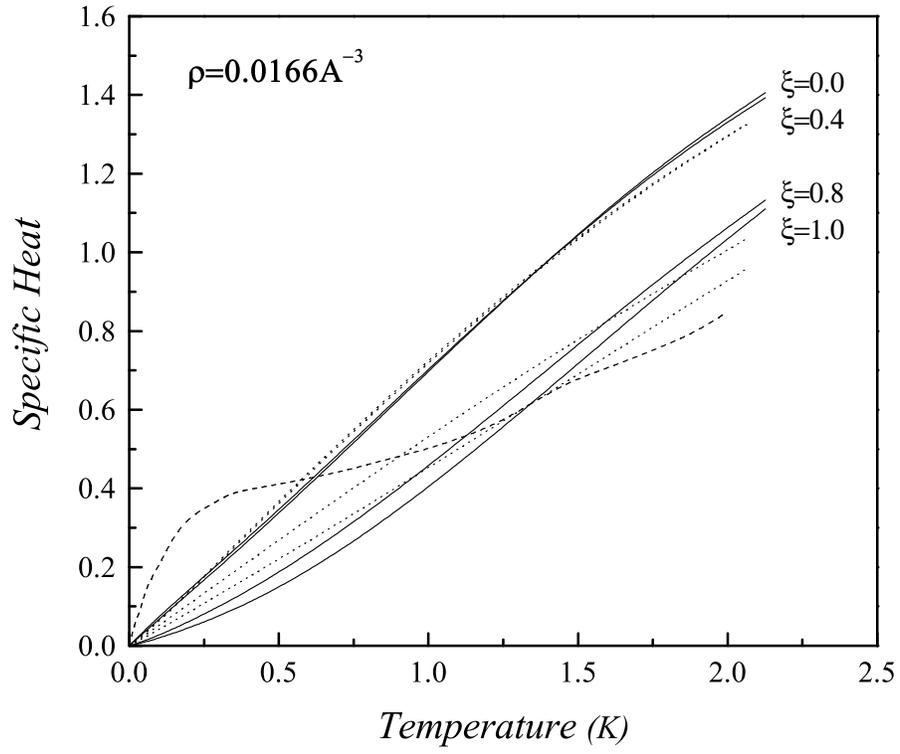}
 \caption{ As Fig. \ref{FigFT}, but
for the specific heat.} \label{FigCVT}
\end{figure}

In Figs. \ref{FigFT}-\ref{FigCVT}, we have shown the free energy,
entropy and the specific heat of liquid  $\mathrm{^3He}$ as a
function of temperature for different values of $\xi$ at
$\rho=0.0166 A^{-3}$. In these figures, the experimental results
\cite{Wi} for the unpolarized liquid $\mathrm{^3He}$ are also
shown for comparison. Fig. \ref{FigFT} indicates a good agreement
between our results for the free energy with the Lennard-Jones
potential and the experimental results. From Figs.
\ref{FigFT}-\ref{FigCVT}, it is seen that the free energy (entropy
and specific heat) decreases (increase) by increasing temperature.
In addition, we see that for the free energy, entropy and the
specific heat of liquid $\mathrm{^3He}$, the differences between
fully polarized and unpolarized cases increase by increasing
temperature.

\begin{figure}
\includegraphics[height=2.7in]{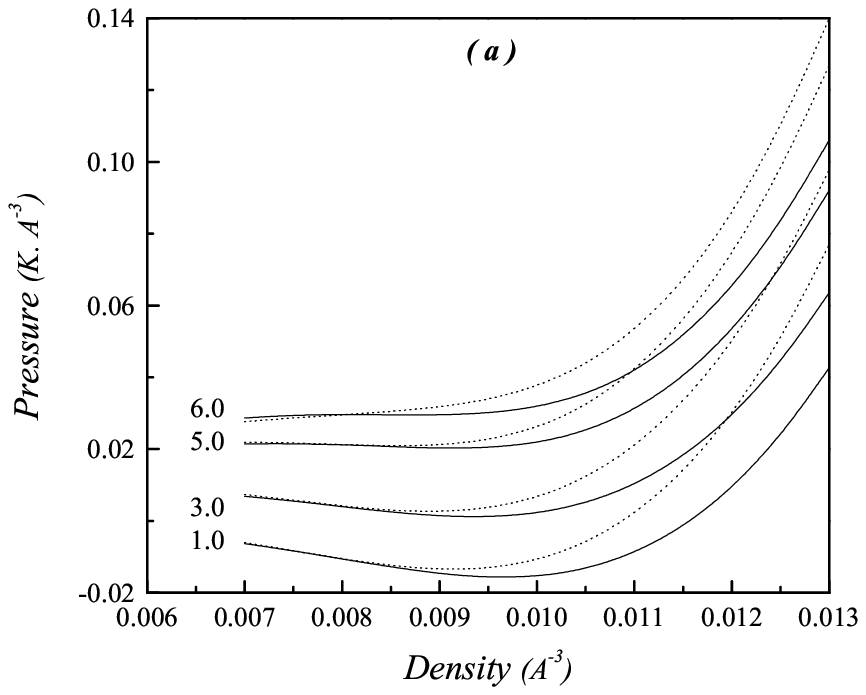}
\includegraphics[height=2.7in]{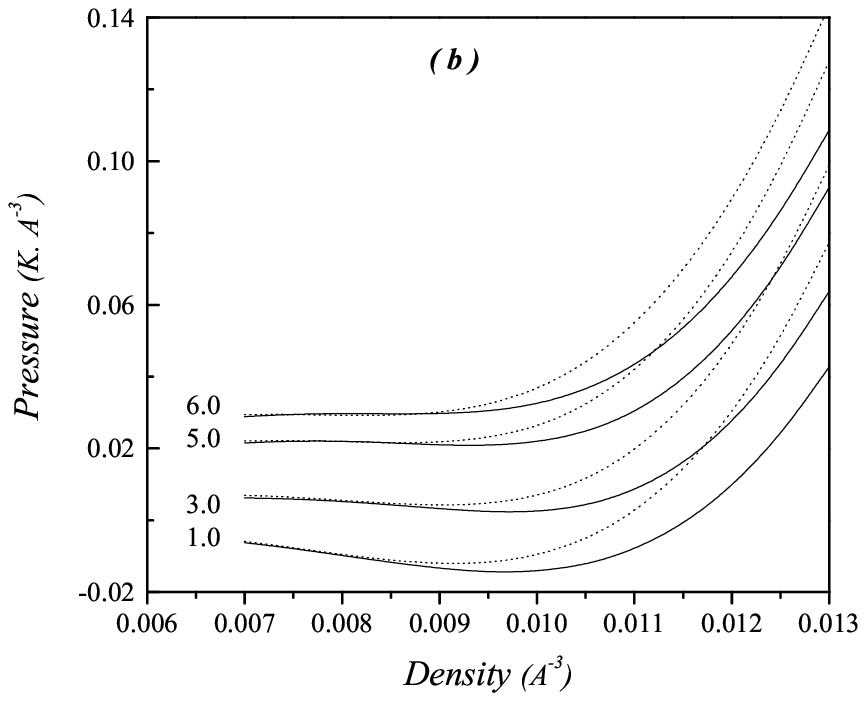}
\includegraphics[height=2.7in]{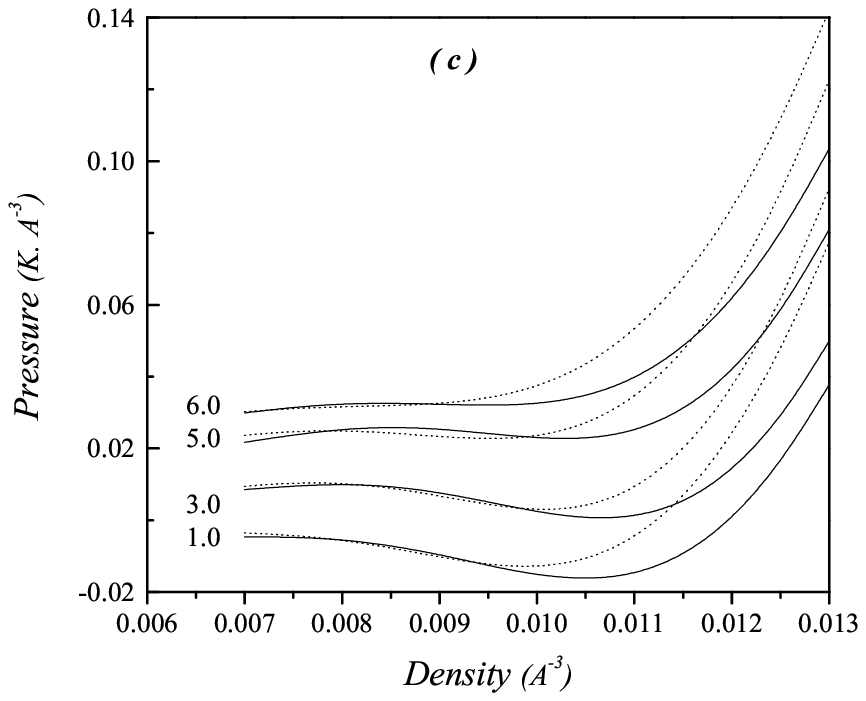}
\includegraphics[height=2.7in]{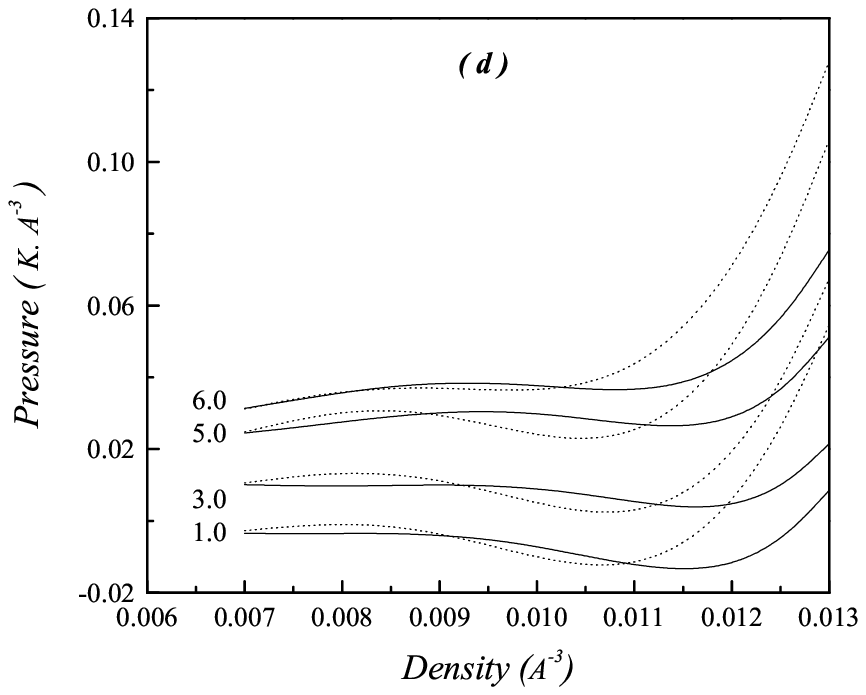}
 \caption{The pressure versus density at different values of temperature
($T=1.0, 3.0, 5.0 , 6.0 K$) with the Leonard-Jones (full curves)
and Aziz (dotted curves) potentials
 for  $\xi=0.0$ (a), $\xi=0.4$ (b), $\xi=0.8$ (c) and $\xi=1.0$
(d).
  } \label{pressure}
\end{figure}

 The isothermal pressure is obtained from the
free energy using the following equation,
\begin{equation}
P(\rho,T)=\rho^{2}\frac{\partial{F(\rho,T)}}{\partial{\rho}}.
\end{equation}
In Fig. \ref{pressure}, we have plotted the isothermal pressure as
a function of density for the various temperatures and $\xi$. It
is seen that the equation of state with the Aziz potential is
stiffer than that of the Lennard-Jones potential. This figure
shows that at low temperatures we have a liquid-gas phase
equilibrium for the liquid $\mathrm{^3He}$.  We know that at the
critical temperature $(T_{c})$ there is no liquid-gas phase
equilibrium. Our results of the critical temperature for different
values of $\xi$ are presented in Table \ref{Tab3}. We see that the
critical temperature increases by increasing $\xi$. For all $\xi$
it is seen that the $(T_{c})$ with the Aziz potential is lower
than that of the Leonard-Jones potential.

For liquid $\mathrm{^{3}He}$, the saturation incompressibility at
saturation density is given by,
\begin{equation}
K_{0}(T)=9\left(\rho^{2}\frac{\partial^{2}F(\rho,T)}
{\partial{\rho^{2}}}\right)_{\rho_{0}}.
\end{equation}
 In Table \ref{Tab2}, our calculated values of the
 saturation incompressibility
 are presented for different values of $\xi$
and $T$. It is seen that for each value of $\xi$, the
incompressibility decreases by increasing temperature and
 at each temperature, it increases by increasing $\xi$.
 From Table \ref{Tab2} we see that for all values of spin asymmetry parameter
 our results for the $K_{0}$
 with the Aziz potential are lower than those of the Leonard-Jones
 potential.

\section{Summary and Conclusion\label{con}}
We have considered a system consisting of the Helium atoms
($^{3}\mathrm {He}$) with an asymmetrical spin configuration. For
this system, we have calculated some of the thermodynamic
properties using the lowest order constrained variational method
with the Lennard-Jones and Aziz potentials. Our calculations lead
to the following conclusions for the liquid $^{3}\mathrm {He}$:

\begin{itemize}
\item At high temperatures, the
correlation function reaches the limiting value for the smaller
values of the interatomic distances.

\item The results of the correlation function with the Lennard-Jones and
Aziz potentials are very similar.

\item The free energy with the the Lennard-Jones and Aziz
potentials are identical for small values of the density.

\item  The free energy, saturation density and
incompressibility decrease  by increasing temperature.

\item  The free energy, saturation density,
incompressibility and critical temperature increase by increasing
spin asymmetry parameter.

\item The entropy, specific heat and flashing temperature
decrease by increasing spin asymmetry parameter.

\item The entropy decreases by increasing density and increases by
increasing temperature.

\item The equation of state with the Aziz potential is stiffer than
that of the Lennard-Jones potential.

\item For all values of spin asymmetry parameter
 our results for the flashing temperature, saturation density,
 critical temperature and saturation incompressibility
 with the Aziz potential are lower than those of the Leonard-Jones
 potential.

\end{itemize}

\acknowledgements{ Financial support from the Shiraz University
research council is gratefully acknowledged.}

\newpage

\begin{table}
\begin{center}
\caption{The flashing temperature ($T_{f}$) for different values
of $\xi$.}\label{Tab1}
\begin{tabular}{ccc}
 \\\hline\hline
 \hspace{2.5cm} Lennard- Jones  & \hspace{1.5cm} Aziz & \hspace{1.0cm}
 \end{tabular}
\begin{tabular}{ccc}
\hline\hline
 \hspace{1cm} $\xi$\hspace{1cm} &\hspace{1cm} $T_{f}(K)$\hspace{1cm}
 &\hspace{1cm} $T_{f}(K)$\hspace{1cm}  \\ \hline
   0.0 &2.8&2.6 \\
   0.4 &2.7&2.5\\
   0.8 &2.6&2.3\\
   1.0 &2.2&1.9\\
   \hline
\end{tabular}
\end{center}
\end{table}

\newpage

\begin{table}
\begin{center}
\caption{The saturation density ($\rho_{0}$) and the saturation
incompressibility ($K_{0}$) of liquid $\mathrm{^{3}He}$ for
different values of $\xi$ and temperature.}\label{Tab2}
\begin{tabular}{ccc}
 \\\hline\hline
 \hspace{4.5cm} Lenard- Jones  & \hspace{3.4cm} Aziz & \hspace{1.0cm}
 \end{tabular}
\begin{tabular}{cccccc}
 \hline\hline
  $\xi$  & \hspace{.8cm} $T (K)$
 &\hspace{.8cm} $\rho_{0} (A^{-3})$
 & \hspace{.8cm} $K_{0} (K)$&\hspace{.8cm} $\rho_{0} (A^{-3})$
 & \hspace{.8cm} $K_{0} (K)$   \\
\hline
0.0 &0.5 & 0.0117&185&0.011&174\\
0.0 &1.0 & 0.0116&163&0.0109&157\\
0.0 &1.5 & 0.0113&135&0.0106&121\\
 \hline
   0.4 & 0.5&0.0117&187&0.011&183\\
   0.4 & 1.0&0.0116&172&0.0109&162\\
   0.4 & 1.5&0.0114 &145&0.0107&132\\
     \hline
   0.8 &0.5 & 0.0120&237&0.0113&212\\
   0.8 & 1.0& 0.0120&232&0.0113&206\\
0.8 & 1.5& 0.0119&203&0.0112&179\\
    \hline
    1.0 & 0.5&0.0128&275&0.0119&260\\
     1.0 & 1.0&0.0128&263&0.0119&246\\
     1.0 & 1.5&0.0127&212&0.0118&204\\
   \hline
\end{tabular}
\end{center}
\end{table}

\newpage

\begin{table}
\begin{center}
\caption{The critical temperature ($T_{c}$) for different values
of $\xi$.}\label{Tab3}
\begin{tabular}{ccc}
 \\\hline\hline
 \hspace{2.5cm} Lennard- Jones  & \hspace{1.5cm} Aziz & \hspace{1.0cm}
 \end{tabular}
\begin{tabular}{ccc}
\hline\hline
 \hspace{1cm} $\xi$\hspace{1cm} &\hspace{1cm} $T_{c}(K)$\hspace{1cm}
 &\hspace{1cm} $T_{c}(K)$\hspace{1cm}  \\ \hline
   0.0 &5.3&5.1 \\
   0.4 &5.5&5.3\\
   0.8 &6.2&5.8\\
   1.0 &6.5&6.1\\
   \hline
\end{tabular}
\end{center}
\end{table}

\newpage

\newpage

\end{document}